# Tunable Kondo Effect in Graphene with Defects

Jian-Hao Chen, W. G. Cullen, E. D. Williams, and M. S. Fuhrer

*Materials Research Science and Engineering Center and Center for Nanophysics and Advanced Materials, Department of Physics, University of Maryland, College Park, MD 20742 USA*

**Graphene is a model system for the study of electrons confined to a strictly two-dimensional layer[1] and a large number of electronic phenomena have been demonstrated in graphene, from the fractional[2, 3] quantum Hall effect to superconductivity[4]. However, the coupling of conduction electrons to local magnetic moments[5, 6], a central problem of condensed matter physics, has not been realized in graphene, and, given carbon's lack of *d* or *f* electrons, magnetism in graphene would seem unlikely. Nonetheless, magnetism in graphitic carbon in the absence of transition-metal elements has been reported[7-10], with explanations ranging from lattice defects[11] to edge structures[12, 13] to negative curvature regions of the graphene sheet[14]. Recent experiments suggest that correlated defects in highly-ordered pyrolytic graphite (HOPG) induced by proton irradiation[9] or native to grain boundaries[7], can give rise to ferromagnetism. Here we show that point defects (vacancies) in graphene[15] are local moments which interact strongly with the conduction electrons through the Kondo effect[6, 16-18] providing strong evidence that defects in graphene are indeed magnetic. The Kondo temperature $T_K$ is tunable with carrier density from 30-90 K; the high $T_K$ is a direct consequence of strong coupling of defects to conduction electrons in a Dirac material[18]. The results indicate that defect engineering in graphene could be used to generate and control carrier-mediated magnetism, and realize all-carbon**

**spintronic devices. Furthermore, graphene should be an ideal system in which to probe Kondo physics in a widely tunable electron system.**

We previously reported the resistivity of graphene with vacancies induced by ion irradiation in ultra-high vacuum (UHV)[15]. Here we present a detailed study of the gate voltage ($V_g$) and temperature ($T$) dependence of the resistivity $\rho(V_g, T)$ in similar graphene with vacancies over a wider temperature range 300 mK < T < 290K. Apart from weak-localization (WL) corrections[19, 20], we find that $\rho(V_g,T)$ is explained by a temperature-independent contribution $\rho_c(V_g)$ due to non-magnetic disorder plus a temperature-dependent contribution $\rho_K(V_g,T)$, not present in as-prepared graphene[15], which follows the universal temperature dependence expected for Kondo scattering from a localized ½-spin with a single scaling parameter $T_K$.

Graphene with vacancies is prepared as described in Ref. 15. After irradiation, the device was annealed overnight at 490 K in UHV, and then exposed to air while transferring to a $^3$He sample-in-vacuum cryostat. Figure 1a shows $\rho(V_g)$ measured at 17K for a graphene device before irradiation, immediately after irradiation, and measured at 300 mK after annealing and transfer to the $^3$He cryostat. $V_g$ is applied to the Si substrate to tune the carrier density $n = c_g V_g/e$, where $c_g = 1.15 \times 10^{-8}$ F/cm$^2$ is the gate capacitance, and $e$ the elementary charge. The mobility of the device is approximately 4000, 300, and 2000 cm$^2$/Vs, respectively, for these three measurements; the conductivity and mobility recover significantly after annealing and air exposure, consistent with our previous study[15]. From the post-annealing mobility we estimate that this device has a defect density of approximately $3 \times 10^{11}$ cm$^{-2}$, although greater understanding of the effects of annealing and ambient exposure on vacancies in graphene is needed. See Supplementary Information for the calculation of defect density and also Raman spectra of the device before and after irradiation.

Figure 1b shows the perpendicular magnetic field dependence of the resistivity $\rho(B)$ of the irradiated sample at $T$ = 300 mK at several different gate voltages. Negative magnetoresistance is observed at small $B$ indicating the dominance of weak localization (WL) arising from intervalley scattering due to lattice defects[19, 20]. Figure 1c shows a detail of the magnetoresistance at small $B$, at 300 mK and at $V_g - V_{g,min}$ = -65 volts. Shubnikov–de Haas (SdH) oscillations appear at high B field. In order to measure the resistivity without WL and SdH corrections, the WL contribution is suppressed by application of $B$ = 1 T in further measurements. For $|V_g - V_{g,min}| < 5$ V, SdH corrections may affect the data slightly at 1 T. However, as shown below, the $\rho(T)$ behavior for $|V_g - V_{g,min}| > 5$ V and $|V_g - V_{g,min}| < 5$ V show no qualitative differences.

Figure 2a shows the temperature-dependent resistivity $\rho(T)$ of the irradiated graphene measured at several different gate voltages at $B$ = 1 T. Positive slopes, $d\rho/dT > 0$, are seen in $\rho(T)$ from room temperature to about 200 K for $V_g$ not too near $V_{g,min}$ indicating phonon contributions[21]; between ~10 K to ~100 K, we find $d\rho/dT < 0$ and the resistivity increases logarithmically with decreasing temperature at all $V_g$. At low temperature the resistivity at all $V_g$ saturates ($d\rho/dT \rightarrow 0$), indicating that there is no disorder-induced metal to insulator transition (MIT)[18] or opening of a band gap[22].

In metallic systems where localized magnetic moments couple anti-ferromagnetically to the conduction electrons, spin-flip scattering gives rise to an anomalous component of the resistivity $\rho_K(T)$ which is characterized by a Kondo temperature $T_K$ (Ref. 6, 23). For $T \approx T_K$, $\rho_K(T)$ is approximately logarithmic in $T$ (similar behavior is observed *in situ* in UHV before ambient exposure; see Supplementary Information). In principle interaction effects in the presence of disorder could also lead to logarithmic ρ($T$) even at high magnetic field (the Altshuler-Aronov effect); however this would also lead to similar corrections to the Hall resistivity, which are not observed (see Supplementary Information). For $T << T_K$, the conduction electrons screen the spins of the local moments and the resistivity saturates, with a negative correction proportional to $T^2$ (Ref. 24). To compare the observed data in graphene with vacancies to theories of the Kondo effect, we model the temperature-dependent resistivity in the low temperature regime and the intermediate temperature regime (region of maximum logarithmic slope, roughly between 10 K and 100 K) as, respectively,

$$\rho(V_g,T) = \rho_{c1}(V_g) + \rho_{K,0}(V_g)\left(1 - \left(\frac{\pi}{2}\right)^4 \left(\frac{T}{T_K(V_g)}\right)^2\right) \quad (1)$$

$$\rho(V_g,T) = \rho_{c2}(V_g) + \frac{\rho_{K,0}(V_g)}{2}\left(1 - 0.470\,\mathrm{Ln}\left(\frac{T}{T_K(V_g)}\right)\right) \quad (2)$$

where $\rho_{K,0}$ is the Kondo resistivity at zero temperature, $\rho_{c1}$ and $\rho_{c2}$ the non-temperature-dependent part of the resistivity, presumably from impurity scattering that does not involve the spin degree of freedom[25, 26]. The numerical factors in Equations 1 and 2 are from theory of the spin-½ Kondo effect[23]. Since ρ($V_g$, $T$ = 0) is known, there are three degrees of freedom in the equations above at each $V_g$: $\rho_{c1}$, $\rho_{c2}$ and $T_K$; if $\rho_K(T)$ follows the universal Kondo form then $\rho_{c1} = \rho_{c2}$. We keep $\rho_{c1}$ and $\rho_{c2}$ as independent parameters to test the internal consistency of the model. Least square fits to the Equation 1 and 2 are carried out on $\rho(V_g,T)$ in the low and intermediate temperature ranges respectively (see Supplementary Information for details).

Using the extracted parameters, we can scale the $\rho(T)$ curves at different $V_g$ and compare them to the universal Kondo behavior[23, 27]. Figure 2b shows the normalized Kondo resistivity $(\rho - \rho_{c1})/\rho_{K,0}$ vs. $T/T_K$ and the universal Kondo behavior from numerical renormalization group calculations (NRG)[23]. From Figure 2b one can find that: 1) all the experimental curves collapse to a single functional form for 300mK < $T$ < ~3$T_K$ and 2) the functional form matches well the universal Kondo behavior from NRG calculations. At higher temperature (T > 200K), phonon contributions become important[21] and the observed positive deviations from the NRG calculations are expected. However, at the lowest gate voltages, the deviation is negative, possibly due to thermal activation of carriers.

Now we discuss the gate voltage dependence of the extracted parameters, $\rho_{c1}$, $\rho_{c2}$, $\rho_{K,0}$ and $T_K$. Figure 3a shows $\rho_{c1}$ and $\rho_{c2}$ vs. $V_g$, which peak around the actual minimum conductivity gate voltage $V_{g,min} \approx 5.3$ volts. We find that $\rho_{c1}$ and $\rho_{c2}$ are practically identical, which indicates that the logarithmic divergence and $T^2$ saturation of the resistivity indeed arise from the same effect (the Kondo effect). From now on we use $\rho_{c1}$ for the non-Kondo resistivity and label it as $\rho_c$. Figure 3b shows the non-Kondo conductivity $G_c = 1/\rho_c$ as a function of $V_g$, which has a similar gate voltage dependence as an as-prepared graphene sample. i.e. linear $G(V_g)$ at high $V_g$, and a minimum $G$ of a few $e^2/h$ where $h$ is Planck's constant. It is worth noting that the minimum non-Kondo conductivity $G_{c,min}$= 6.9 $e^2/h$ is the same as the minimum conductivity of the pristine sample before irradiation (see Fig. 1a).

Figure 3c shows the Kondo resistivity $\rho_{K,0} = \rho(T = 0) - \rho_c$ vs. $V_g$, which also peaks around the $V_{g,min}$, and decreases rapidly with increasing $|V_g - V_{g,min}|$, and Figure 3d shows $G_{K,0} = 1/\rho_{K,0}$ vs. $V_g$. In the low-temperature limit (saturated resistivity), we expect $G_{K,0} \approx \frac{\pi e^2}{h}\frac{n}{n_{imp}}$ (Ref. 17).

However, $G_{K,0}$ is 3-10 times larger than expected for $n_{\mathrm{imp}} = 3 \times 10^{11}$ cm$^{-2}$, and varies more rapidly; the red solid line is a power law fit to $G_{K,0}(V_g)$ that yields $G_{K,0} \sim A + BV_g^{\alpha}$ with $\alpha = 2.1 \pm 0.1$ for electron conduction and $\alpha = 2.2 \pm 0.2$ for hole conduction.

Figure 4 shows the gate voltage dependence of the Kondo temperature $T_K$. $T_K$ is of order 50 K, which indicates strong coupling between the localized magnetic moment and the conduction electrons. Moreover, $T_K$ is tunable by gate voltage, with a minimum of about 30K near $V_{g,min}$ and maxima close to 90K at $V_g$ - $V_{g,min} \approx$ -20 volts (hole conduction) and close to 70K at $V_g$ - $V_{g,min} \approx$ 25 volts (electron conduction). At higher gate voltages ($|V_g - V_{g,min}| >$ 25 volts), $T_K$ decreases slightly with gate voltage, although the experimental error becomes large at large $V_{\mathrm{g}}$ as the Kondo resistivity become very small.

To understand the gate voltage dependence and the large Kondo temperature, we can use the rough estimate for its physical origins:

$$k_B T_K \approx D e^{1/J\rho(E_F)} \tag{3}$$

where $k_{\mathrm{B}}$ is Boltzmann's constant, $D$ is the electronic bandwidth (~10 eV for π electrons in graphene), $J$ the coupling constant (which must be negative for the Kondo effect to exist), and $\rho(E_F) = 8\pi E_F / \left(hv_F\right)^2 = 4\sqrt{\pi n} / hv_F$ is the density of states of graphene at the Fermi energy. The decrease of $T_K$ with decreasing carrier density for -20 V < $V_g$ - $V_{g,min}$ < +25 V is qualitatively consistent with Equation 3 given the decrease in $\rho(E_F)$ with $V_{\mathrm{g}}$. The fact that $T_K$ does not decrease exponentially with $\rho(E_F) \sim n^{1/2} \sim V_{\mathrm{g}}^{1/2}$ indicates electron and hole puddles[28] play an important role in producing a finite effective $\rho(E_F)$ and hence finite $T_{\mathrm{K}}$ at all $V_{\mathrm{g}}$.

We estimate the density of states $\rho(E_F)$ varies over a range of 0.003 – 0.01 $eV^{-1}$ per atom from $0 < V_g - V_{g,min} < 50$ V. A Kondo temperature of $T_K$ = 30 – 90 K corresponds to an apparently unphysically large value of $J$ = 10 – 40 eV using Equation 3. However, it has been pointed out that the dimensionless parameter which determines the strength of the Kondo effect, $J\rho(E_F)$ in Equation 3, is proportional to the sine of the scattering phase shift for the defect, which for the large scalar potentials induced by defects in the graphene plane may be of order unity[18], consistent with the very large value of $J$ determined here. Thus the unique properties of the Dirac Hamiltonian allow a robust $T_K$ even for a low density of states. It is also probable that in disordered graphene with potential fluctuations, the resistivity probes only the few defects with the largest $T_K$; this is consistent with $G_{K,0}$ larger than expected in the low-temperature limit (Fig. 3d). A complete theory of the Kondo effect in graphene will likely require both a microscopic understanding of the defect and its interaction with conduction electrons, as well as an effective medium theory of Kondo scattering in the presence of potential variations in disordered graphene.

In conclusion, lattice defects (vacancies) in graphene created by irradiation with low energy $He^+$ are found to have local magnetic moments. Such moments couple strongly to conduction electrons in graphene, resulting in the Kondo effect with a gate-tunable Kondo temperature $T_K$ ranging from 30 K – 90 K. The high $T_K$ in graphene with its small density of states is a unique consequence of defect scattering in a Dirac system[18]. Defect engineering thus provides a powerful route to introduce and control magnetism in carbon nanostructures such as graphene and carbon nanotubes. The observation of Kondo scattering from defects in graphene may also explain the anomalous short spin lifetimes observed in graphene spin valves[29]; since a small native concentration of defects could be present in these (and perhaps all) graphene devices[30].

## Acknowledgements

This work has been supported by NSF-UMD-MRSEC grant DMR 05-20471 (JHC, WGC, EDW, MSF) and the US ONR grant N000140610882 (WGC, EDW, MSF). The MRSEC SEFs were used in this work. Infrastructure support has also been provided by the UMD NanoCenter and CNAM. We would also like to thank David Goldhaber-Gordon, Sankar Das Sarma, Enrico Rossi, Euyheon Hwang and Jun Zhu for useful discussions.

## References

1. Geim, A. K. Graphene: Status and Prospects. Science 324, 1530-1534 (2009).
2. Bolotin, K. I., Ghahari, F., Shulman, M. D., Stormer, H. L. & Kim, P. Observation of the fractional quantum Hall effect in graphene. Nature 462, 196-199 (2009).
3. Castro, E. V., López-Sancho, M. P. & Vozmediano, M. A. H. Pinning and switching of magnetic moments in bilayer graphene. New Journal of Physics 11, 095017 (2009).
4. Heersche, H. B., Jarillo-Herrero, P., Oostinga, J. B., Vandersypen, L. M. K. & Morpurgo, A. F. Bipolar supercurrent in graphene. Nature 446, 56-59 (2007).
5. Anderson, P. W. Localized Magnetic States in Metals. Physical Review 124, 41 (1961).
6. Kondo, J. Resistance Minimum in Dilute Magnetic Alloys. Progress of Theoretical Physics 32 37-49 (1964).
7. Cervenka, J., Katsnelson, M. I. & Flipse, C. F. J. Room-temperature ferromagnetism in graphite driven by two-dimensional networks of point defects. Nature Physics 5, 840-844 (2009).
8. Esquinazi, P. et al. Ferromagnetism in oriented graphite samples. Physical Review B 66, 024429 (2002).
9. Esquinazi, P. et al. Induced Magnetic Ordering by Proton Irradiation in Graphite. Physical Review Letters 91, 227201 (2003).
10. Ugeda, M. M., Brihuega, I., Guinea, F. & Gomez-Rodriguez, J. M. Missing Atom as a Source of Carbon Magnetism. Physical Review Letters 104, 096804 (2010).
11. Lehtinen, P. O., Foster, A. S., Ma, Y., Krasheninnikov, A. V. & Nieminen, R. M. Irradiation-Induced Magnetism in Graphite: A Density Functional Study. Physical Review Letters 93, 187202 (2004).
12. Fujita, M., Wakabayashi, K., Nakada, K. & Kusakabe, K. Peculiar Localized State at Zigzag Graphite Edge. Journal of the Physical Society of Japan 65, 1920-1923 (1996).
13. Son, Y. W., Cohen, M. L. & Louie, S. G. Half-metallic graphene nanoribbons. Nature 444, 347-349 (2006).
14. Park, N. et al. Magnetism in All-Carbon Nanostructures with Negative Gaussian Curvature. Physical Review Letters 91, 237204 (2003).
15. Chen, J.-H., Cullen, W. G., Jang, C., Fuhrer, M. S. & Williams, E. D. Defect scattering in graphene. Physical Review Letters 102, 236805 (2009).
16. Sengupta, K. & Baskaran, G. Tuning Kondo physics in graphene with gate voltage. Physical Review B 77, 045417 (2008).

17. Cornaglia, P. S., Usaj, G. & Balseiro, C. A. Localized Spins on Graphene. Physical Review Letters 102, 046801 (2009).
18. Hentschel, M. & Guinea, F. Orthogonality catastrophe and Kondo effect in graphene. Physical Review B 76, 115407 (2007).
19. Morpurgo, A. F. & Guinea, F. Intervalley Scattering, Long-Range Disorder, and Effective Time-Reversal Symmetry Breaking in Graphene. Physical Review Letters 97, 196804 (2006).
20. McCann, E. et al. Weak-Localization Magnetoresistance and Valley Symmetry in Graphene. Physical Review Letters 97, 146805 (2006).
21. Chen, J.-H., Jang, C., Xiao, S., Ishigami, M. & Fuhrer, M. S. Intrinsic and Extrinsic Performance Limits of Graphene Devices on SiO2. Nature Nanotechnology 3, 206 - 209 (2008).
22. Pedersen, T. G. et al. Graphene Antidot Lattices: Designed Defects and Spin Qubits. Physical Review Letters 100, 136804 (2008).
23. Costi, T. A. & et al. Transport coefficients of the Anderson model via the numerical renormalization group. Journal of Physics: Condensed Matter 6, 2519 (1994).
24. Nozières, P. A “fermi-liquid” description of the Kondo problem at low temperatures. Journal of Low Temperature Physics 17, 31-42 (1974).
25. Chen, J.-H. et al. Charged Impurity Scattering in Graphene. Nature Physics 4, 377 (2008).
26. Jang, C. et al. Tuning the effective fine structure constant in graphene: opposing effects of dielectric screening on short- and long-range potential scattering. Physical Review Letters 101, 146805 (2008).
27. Goldhaber-Gordon, D. et al. From the Kondo Regime to the Mixed-Valence Regime in a Single-Electron Transistor. Physical Review Letters 81, 5225 (1998).
28. Rossi, E. & Das Sarma, S. Ground State of Graphene in the Presence of Random Charged Impurities. Physical Review Letters 101, 166803 (2008).
29. Popinciuc, M. et al. Electronic spin transport in graphene field-effect transistors. Physical Review B 80, 214427 (2009).
30. Ni, Z. H. et al. On resonant scatterers as a factor limiting carrier mobility in graphene. Preprint at arXiv:1003.0202 (2010).

**Figure Captions**

**Figure 1 Gate voltage dependent conductivity $\sigma(V_g)$ and magnetoresistance of the graphene sample.** **a**, $\sigma(V_g)$ of the graphene sample before (black solid line) and after (red dashed line) irradiation with 500 eV $He^+$ at a temperature $T$ = 17 K, and after annealing at 490K overnight in UHV and exposure to ambient before cooling to $T$ = 300 mK (blue short-dashed line). Magnetic field $B$ = 0 for all data. The gate voltage of minimum conductivity $V_{g,min}$ = -8 volts, 5 volts, 5.3 volts for pristine, irradiated and annealed sample, respectively. **b**, Magnetoresistance of irradiated and annealed graphene sample for $B$ = 0-8 Tesla at various $V_g$. **c**, normalized detailed magnetoresistance of irradiated and annealed graphene sample from -1.2 Tesla to 1.2 Tesla at $V_g - V_{g,min} \approx -65$ volts.

**Figure 2 Universal Kondo behavior of graphene with defects.** **a**, Temperature-dependent resistivity $\rho(V_g)$ of graphene under perpendicular magnetic field of 1 Tesla, at 12 different gate voltages, with temperature changing from 300mK to ~290 K. **b**, The normalized Kondo part of the resistivity $(\rho-\rho_{c1})/\rho_{K,0}$ versus $T/T_K(V_g)$, where $T_K(V_g)$ is the Kondo temperature at respective gate voltage (see Figure 4). The red line is the expected universal Kondo behavior from numerical renormalization group calculations[23].

**Figure 3 The non-Kondo and Kondo part of the resistivity vs. $V_g$.** **a**, Comparison between the non-Kondo resistivity obtained from fitting $\rho(V_g, T)$ to Eqn. 1 in the low temperature regime ($\rho_{c1}$) and to Eqn. 2 in the intermediate temperature regime ($\rho_{c2}$) at different $V_g$. **b**, non-Kondo conductivity $G_c = 1/\rho_c$ as a function of $V_g$. **c**, The zero temperature Kondo resistivity $\rho_{K,0}$ and **d**, Kondo conductivity $G_{K,0}$ as a function of $V_g$. The red solid line in **d** is a power law fit to $G_{K,0}$,$(V_g)$. The blue dashed line is the expectation for unitary scatterers of concentration $3 \times 10^{11}$ $cm^{-2}$.

**Figure 4 The Gate-tunable Kondo Temperature.** The Kondo temperature $T_K$ of graphene with vacancies as a function of gate voltage $V_g$. as determined from fits to Equations 1 and 2. The error bars represent the ± one standard deviation of $T_K$, calculated using error propagation from the standard deviation of raw fitting parameters (see Supplementary Information).

**Figures**

**Figure 1**

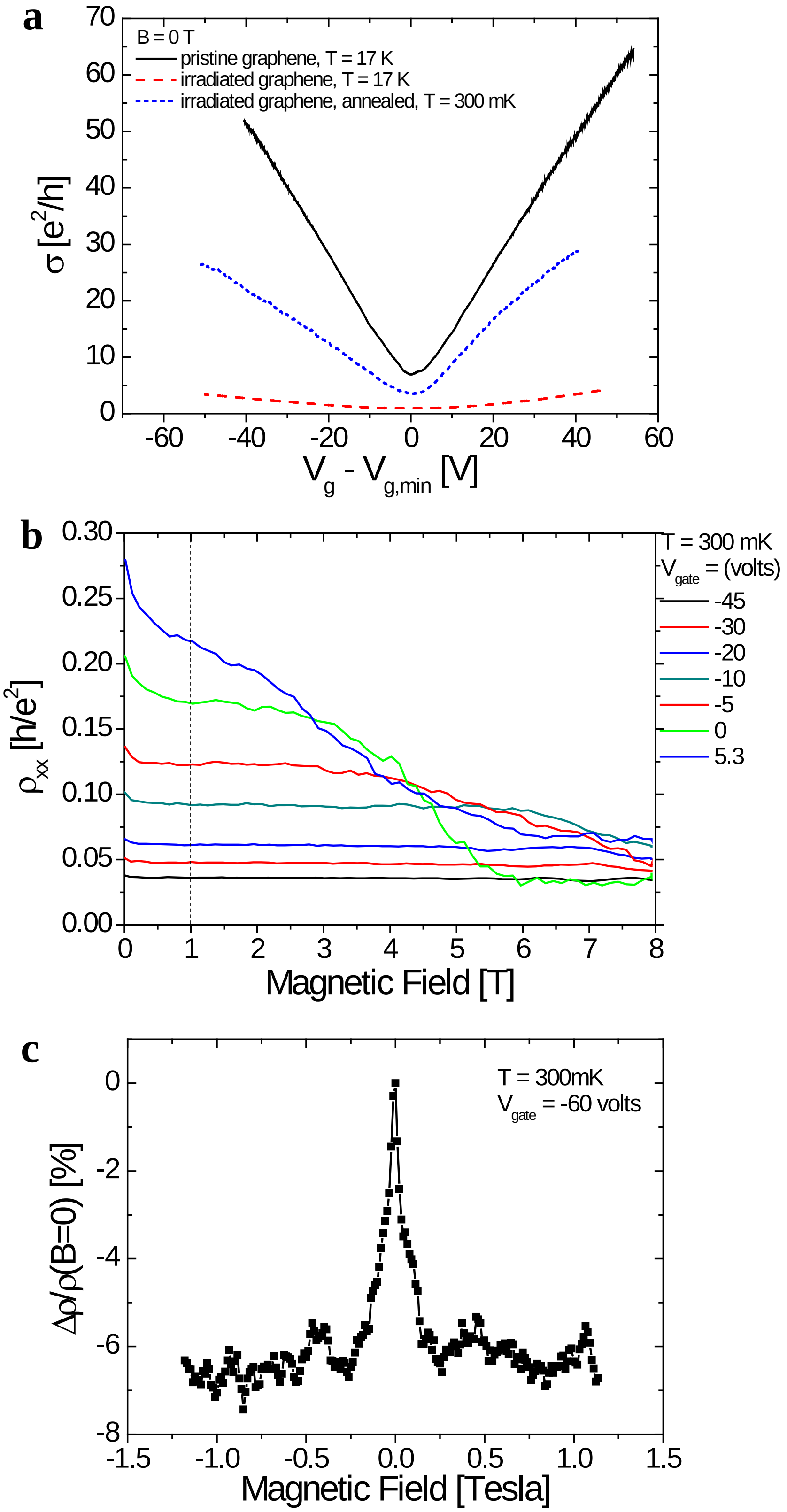

**Figure 2**

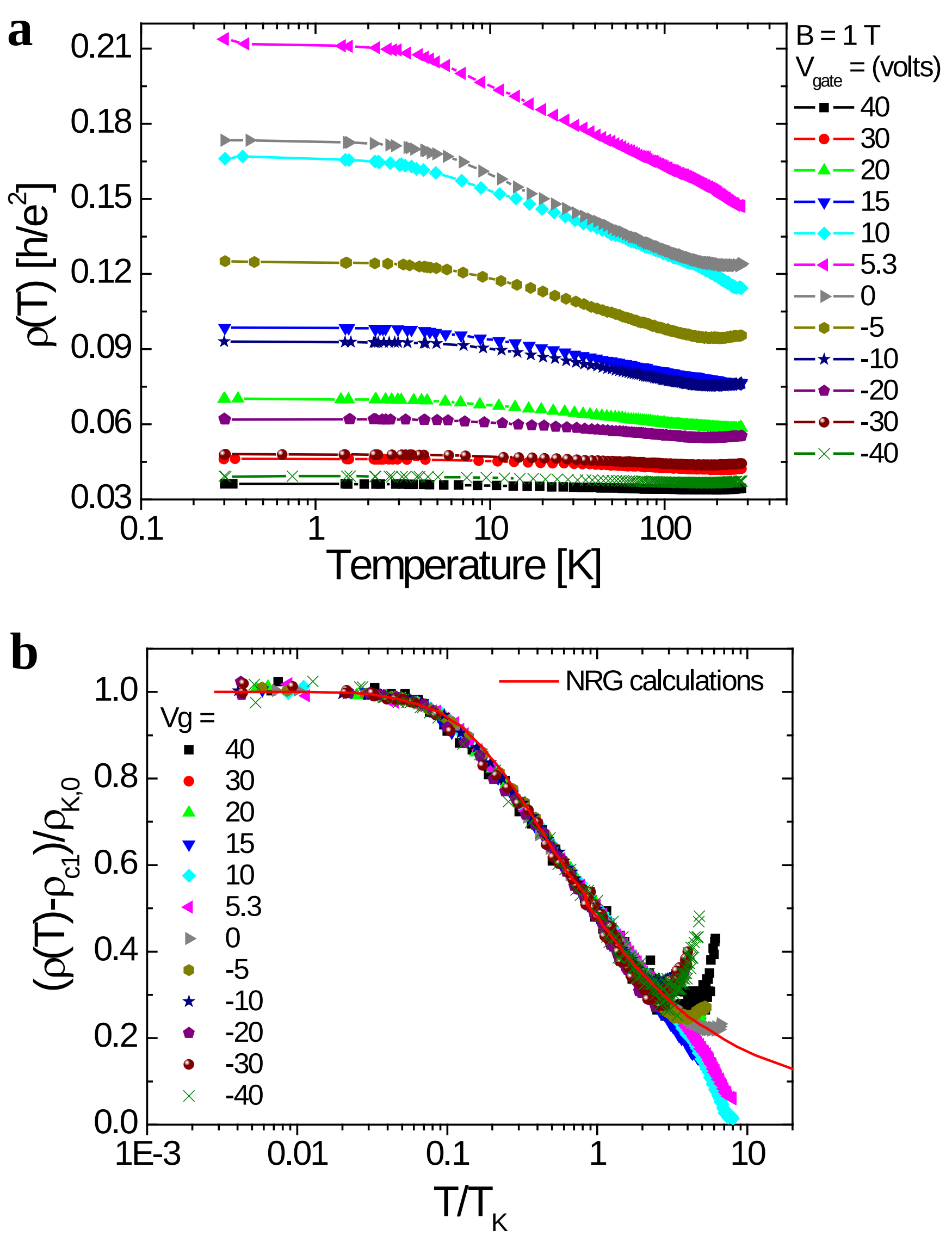

**Figure 3**

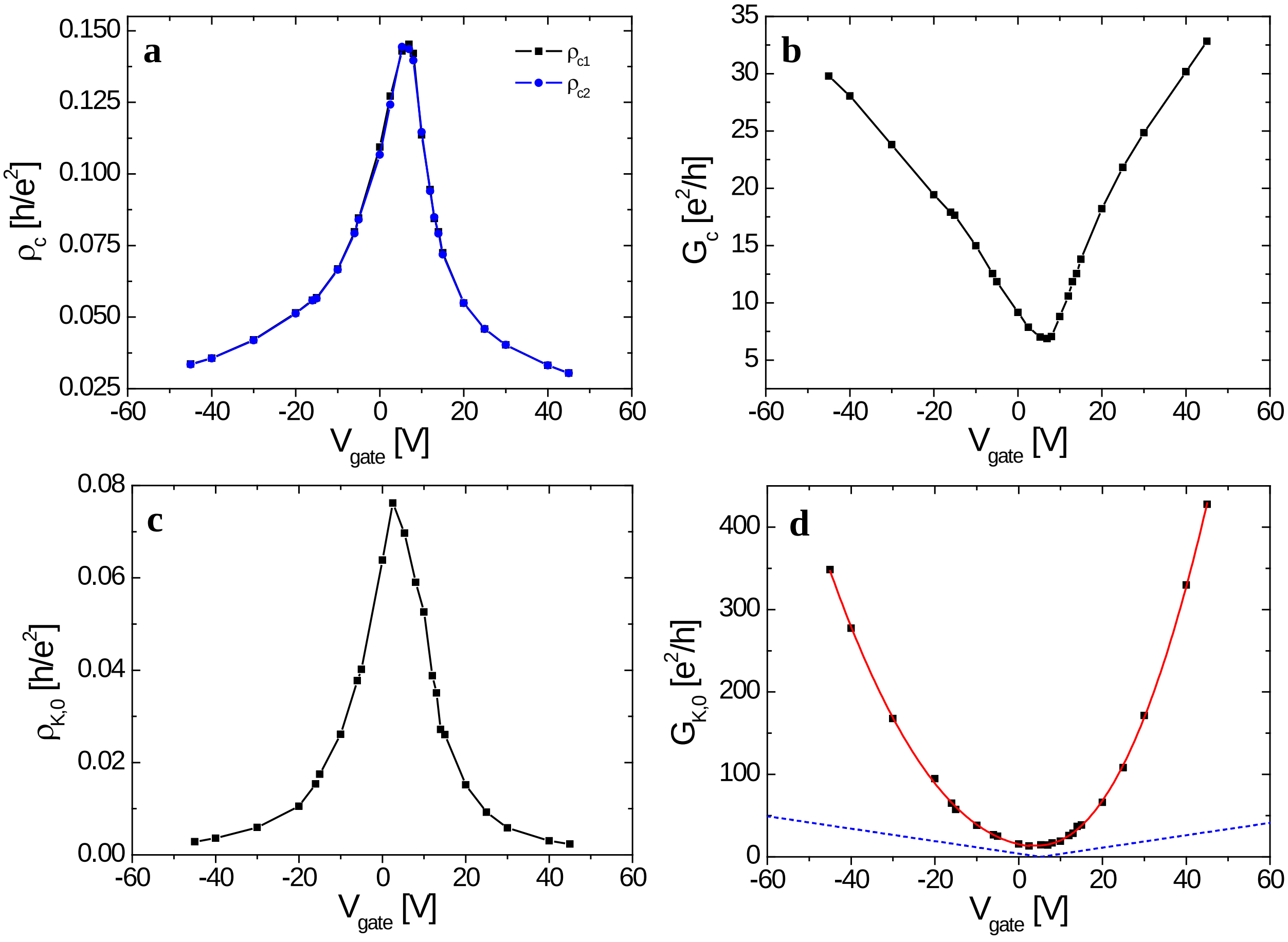

**Figure 4**

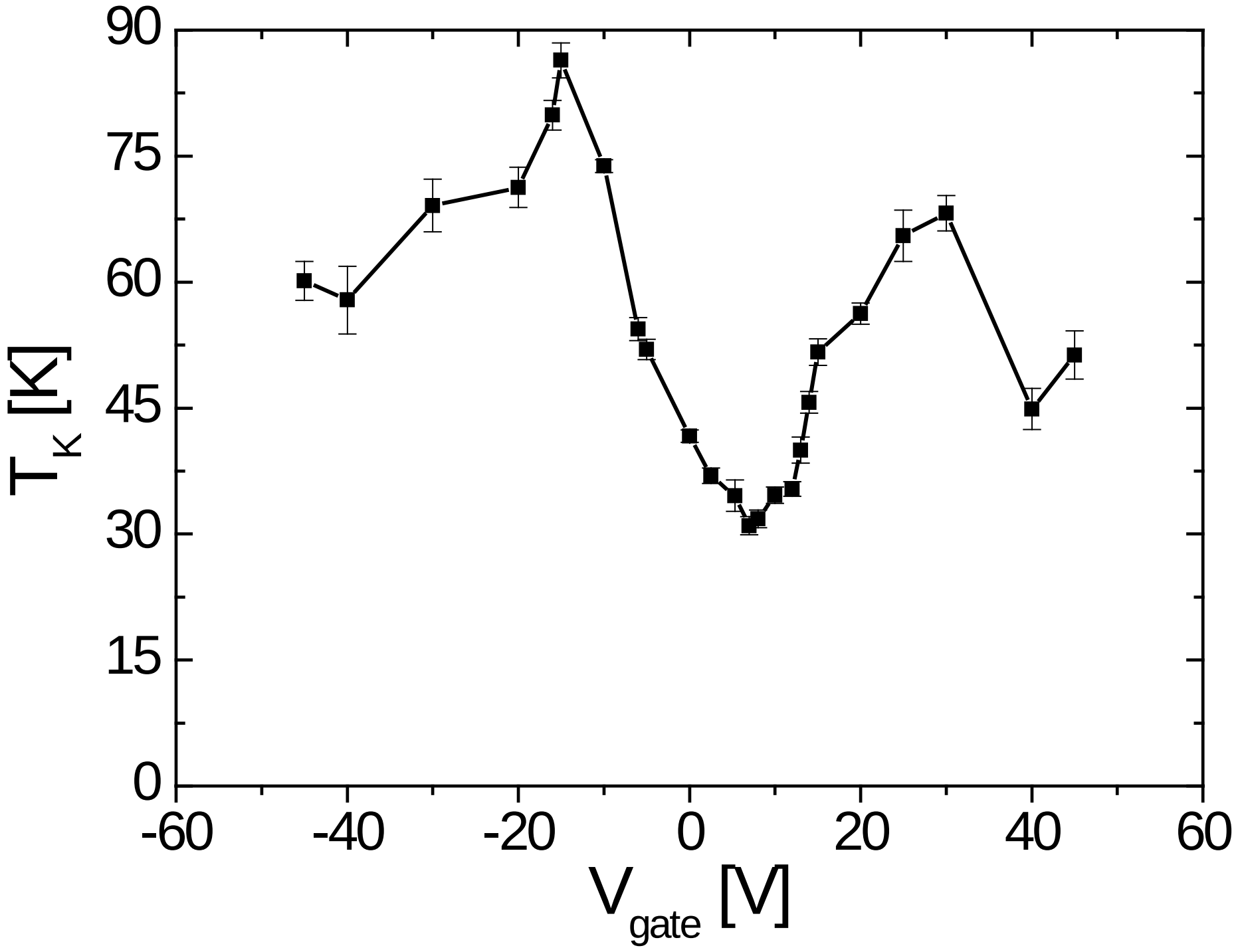

Tunable Kondo Effect in Graphene with Defects

Jian-Hao Chen, W. G. Cullen, E. D. Williams, and M. S. Fuhrer

*Materials Research Science and Engineering Center and Center for Nanophysics and Advanced Materials, Department of Physics, University of Maryland, College Park, MD 20742 USA*

## Supplementary Information

### 1. Estimation of defect density from transport measurement

Because of the charging effect of the insulating substrate (the $SiO_2$ substrate as well as the insulation on the sample holder) under large ion dosages, the number of induced defects is no longer linearly proportional to ion flux and time. Hence we use the proportionality found for smaller ion dosage[1] (presumably under the linear response regime) as well as the mobility before and after irradiation to roughly estimate the effective ion dose.

The total mobility $\mu$ of the sample is given by the combined effect of defects and charged impurities according to Matthiessen's rule as:

$$\frac{1}{\mu} = \frac{1}{\mu_D} + \frac{1}{\mu_C}, \qquad \text{(S1)}$$

where $\mu_D$ is the charge carrier mobility from defect scattering and $\mu_C$ is the mobility from charged impurity scattering. The relation between $\mu_D$ and the defect density $n_D$ is found to be[1]:

$$n_D = \frac{1.2 \times 10^{15} V^{-1} s^{-1}}{\mu_D}. \qquad \text{(S2)}$$

For the graphene sample before irradiation, $n_D \rightarrow 0$ as evident from the absent of the Raman D band, giving $\mu_D \rightarrow \infty$ and $\mu_C = \mu \approx 4000$ cm$^2$/Vs. We assume the mobility from charged impurities remains the same $\mu_C \approx 4000$ cm$^2$/Vs as $He^+$ irradiation does not introduce significant amount charged impurities to the deivce[1]. After irradiation in the UHV chamber, $\mu \approx 300$ cm$^2$/Vs and Equations S1 and S2 give $\mu_D \approx 325$ cm$^2$/Vs and $n_\mathrm{D} \approx 4 \times 10^{12}$ cm$^{-2}$. After annealing under UHV, exposure to ambient, and transfer to the $^3$He cryostat, the low-temperature mobility of the sample $\mu \approx 2000$ cm$^2$/Vs. Assuming that the effect of annealing and ambient exposure is only to remove or passivate some defects, and the remaining defects obey Equation S1, then after annealing $\mu_D \approx 4000$ cm$^2$/Vs and $n_D \approx 3 \times 10^{11}$ cm$^{-2}$. This is a rough estimate, and more work is needed to understand the effect of annealing and ambient exposure on defects in graphene.

## 2. Raman spectra of the graphene device before and after $He^+$ irradiation

Figure S1 shows the Raman spectra, taken under ambient conditions, for the graphene sample studied in the main text before irradiation (a), and after irradiation (b) by 500 eV of $He^+$ at 17 K and subsequent annealing and exposure to ambient. The pristine sample shows a Lorentzian *2D* band characteristic of single layer graphene, and no detectable *D* band. Upon irradiation, the appearance of a very prominent *D* band indicates significant intervalley scattering[2, 3].

**3. *In situ* temperature dependence of irradiated graphene sample**

Temperature dependent resistivity has been measured in UHV for multiple samples after irradiation and overnight annealing without breaking vacuum at a limited temperature range (11K < T < 400 K). Figure S2 shows $\rho(T)$ for one of such graphene sample at zero magnetic field. $V_{g,min}$ = -1 volt for this sample. The logarithmic temperature dependence of the resistivity is consistent with Kondo scattering as observed in the present work.

**4. Absence of Altshuler-Aronov effect in irradiated graphene samples**

The Altshuler-Aronov (AA) effect[4] predicts a logarithmic correction to the resistivity in disordered two dimensional electron systems due to electron-electron interactions, i.e. $\rho_{xx}(T) = \rho_{xx,0} + \delta\rho_{xx}(T)$, where $\delta\rho_{xx}(T) \sim \ln(T)$. The AA effect persists to high magnetic fields, and hence we must consider whether the temperature dependence observed could arise from AA corrections to the resistivity. The AA effect also predicts corrections to the Hall coefficient $R_H = R_{H,0} + \delta R_H(T)$ where $\delta R_H(T)$ is related to $\delta\rho_{xx}(T)$ by

$$\frac{\delta R_H(T)}{R_{H,0}} = 2\frac{\delta\rho_{xx}(T)}{\rho_{xx,0}} \quad . \tag{S3}$$

Figure S3a shows the gate voltage dependence of the resistivity of the same sample (with more irradiation) shown in Figure S2 at various temperatures. Figure S3b shows the gate-voltage dependence of the Hall coefficient for the same sample at the same temperatures. The Hall coefficient is temperature-independent for gate voltages outside of the minimum conductivity region $|V_g - V_{g,min}| > 10$ V, while the resistivity corrections are noticeably large in this region. Therefore we can rule out AA corrections as the source of the logarithmic temperature-dependent resistivity seen in this sample.

**5. Fitting procedure to obtain resistivity scaling parameters**

The fitting of $\rho(V_g,T)$ to the Equation 1 is carried out at 300mK < $T$ < $T_1$, where $T_1$ is the maximum temperature at which the error of the fit is small (i.e. before the $\rho(V_g,T)$ curves start to cross over to the ~log(1/$T$) regime at higher temperature). $T_1$ is found to between 4K (near $V_{g,min}$) to 10K (higher $V_g$). The fitting of $\rho(V_g,T)$ to the Equation 2 is carried out at $T_2$ < T < 70K, where $T_2$ are the smallest temperature where the $\rho(V_g,T)$ ~ log(1/$T$) behavior holds. $T_2$ is essentially bounded by $T_1$, with values between 11K (near $V_{g,min}$) and 28 K (higher $V_g$); the higher bound of the fitting is set to be 70K, to avoid the effects of either phonons (at high $V_g$) or thermal activation (near $V_{g,min}$).

The raw fitting parameters can be written as $a$, $b$, $c$ and $d$, where $\rho(V_g,T) = a + b\times T^2$ (Equation 1) and $\rho(V_g,T) = c + d \times \ln(T)$ (Equation 2), and the standard errors of $a$, $b$, $c$ and $d$ are given by the fits. The maximum standard error (for all $V_g$) of the fitting parameters $\rho_{c1}$, $\rho_{c2}$, $\rho_{K,0}$ and $T_K$ are estimated to be 3%, 5%, 11% and 7%, using standard error propagation formula. The standard error of $T_K$ at each $V_g$ is plotted in Figure 4.

**References**

1. Chen, J.-H., Cullen, W. G., Jang, C., Fuhrer, M. S. & Williams, E. D. Defect scattering in graphene. Physical Review Letters 102, 236805 (2009).
2. Thomsen, C. & Reich, S. Double Resonant Raman Scattering in Graphite. Physical Review Letters 85, 5214 (2000).
3. Narula, R. & Reich, S. Double resonant Raman spectra in graphene and graphite: A two-dimensional explanation of the Raman amplitude. Phys. Rev. B 78, 165422-6 (2008).
4. Altshuler, B. L., Aronov, A. G. & Lee, P. A. Interaction Effects in Disordered Fermi Systems in Two Dimensions. Physical Review Letters 44, 1288 (1980).

## Supplementary Figures

### Figure S1

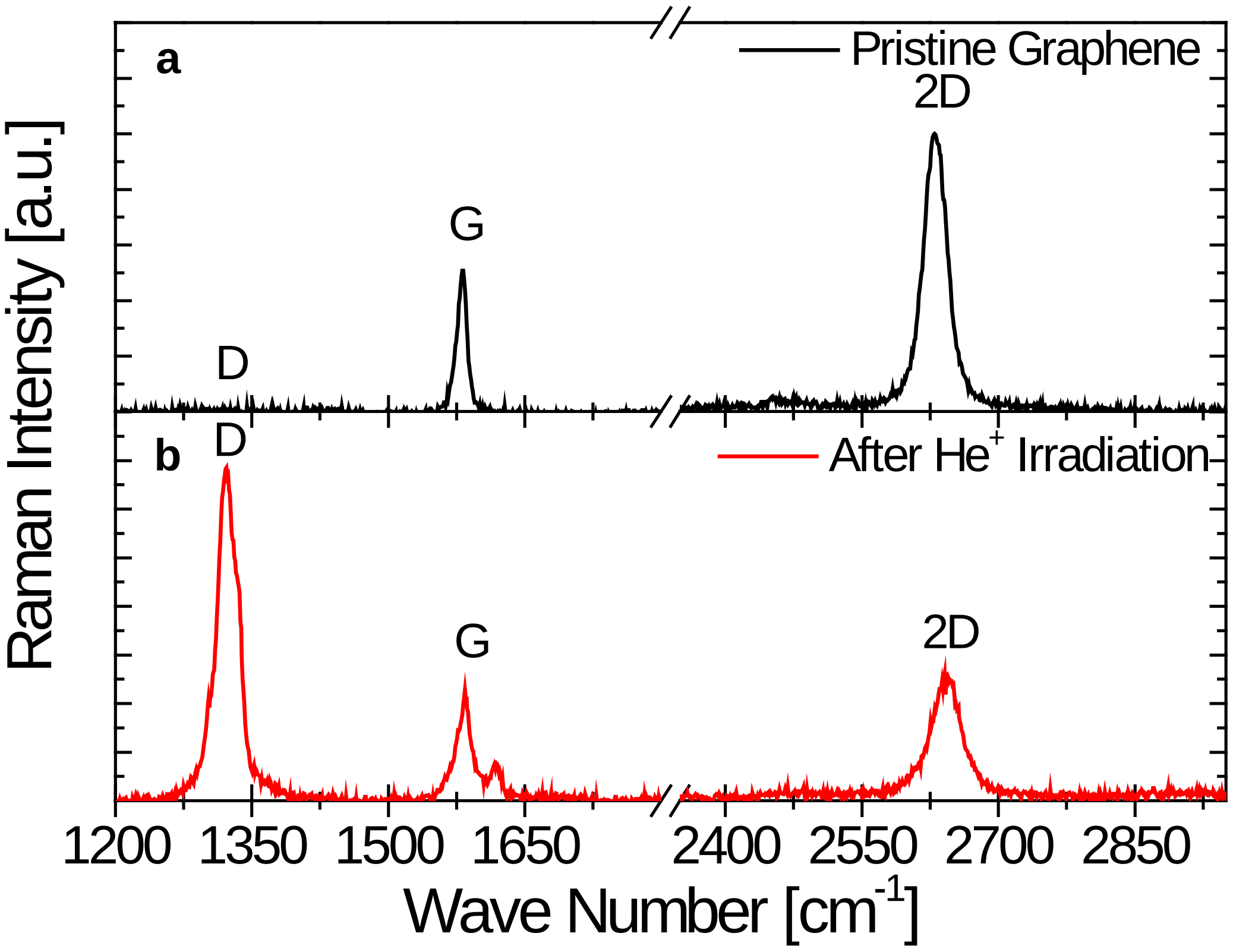

**Figure S1 Raman Spectra of the graphene sample before and after irradiation.** Raman spectra of the graphene sample studied in the main text **a**, before irradiation and **b**, after $He^+$ irradiation and annealing overnight at 490K in UHV. Raman spectra are acquired under ambient conditions before and after the transport measurements using a 633nm laser.

**Figure S2**

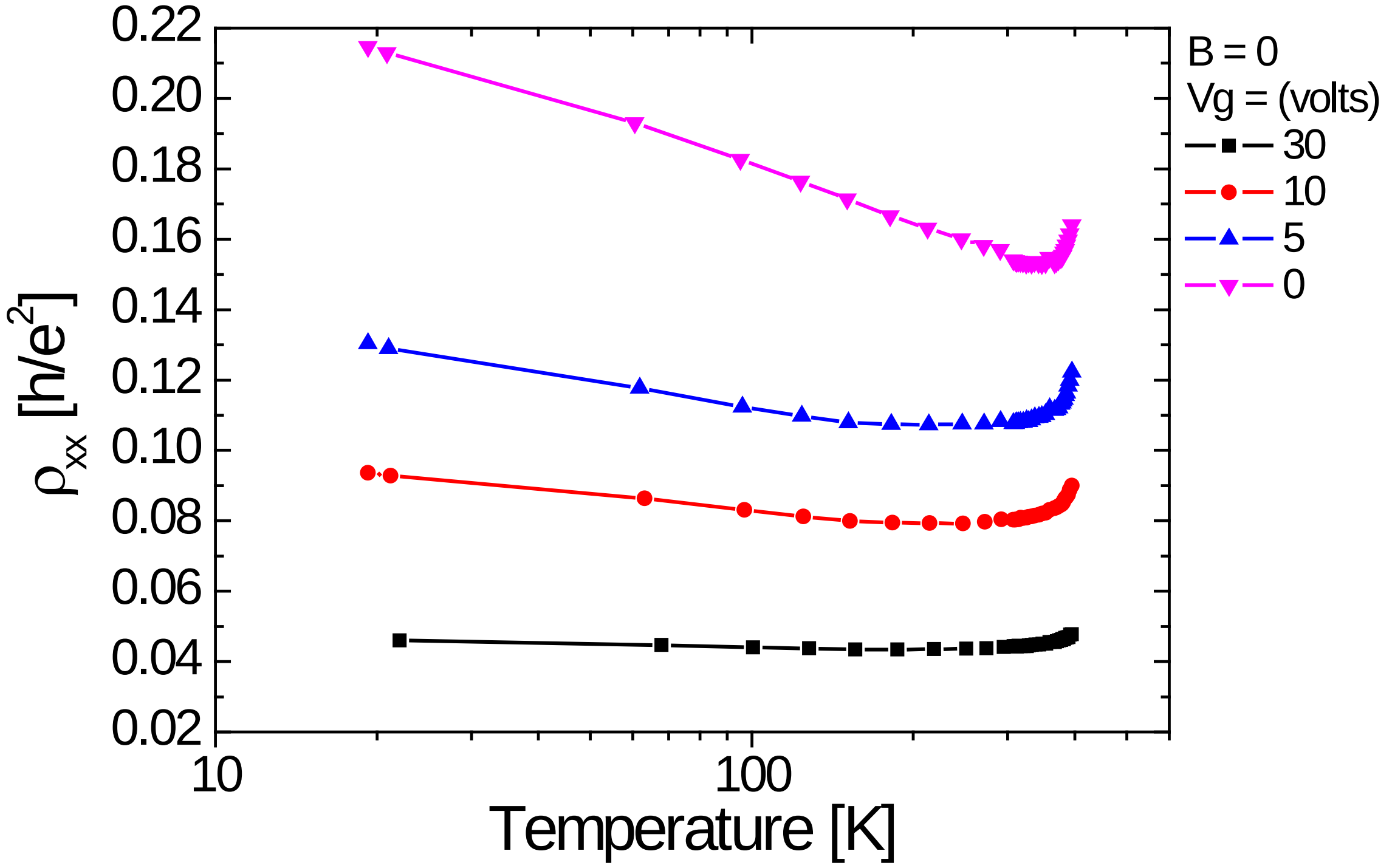

**Figure S2 *In situ* measurement of temperature dependent resistivity of graphene with defects.** $\rho(T)$ for a similar graphene sample to the one discussed in the main text after irradiation with $He^+$ and annealing to 490K overnight at different $V_g$. The measurement is done in the UHV chamber without breaking vacuum.

**Figure S3**

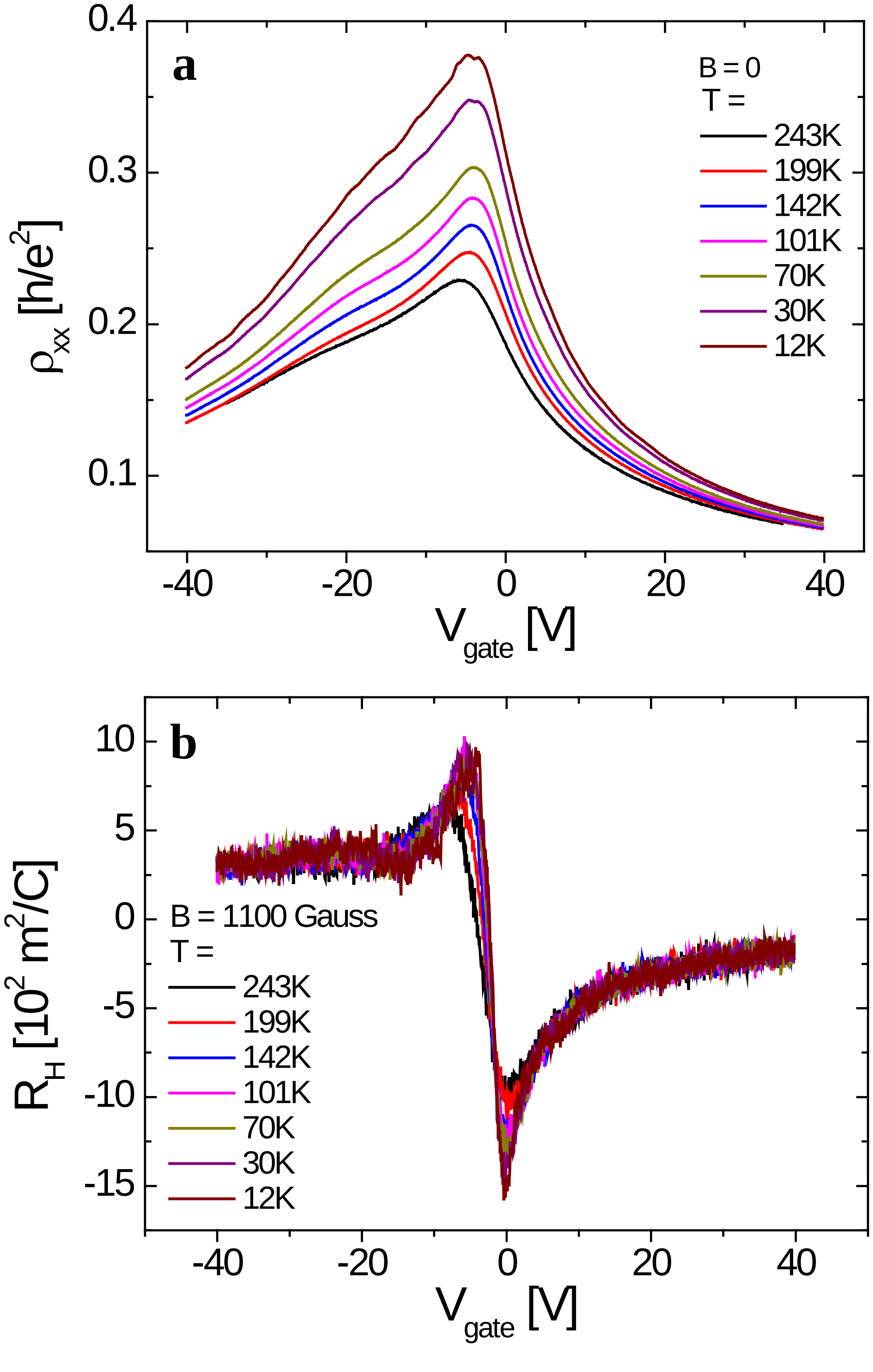

**Figure S3 Temperature dependence of longitudinal resistivity and Hall coefficient.** **a**, Longitudinal resistivity $\rho_{xx}$ versus $V_g$ at various temperatures $T$ for the same sample (with more irradiation) as shown in Figure S2. **b**, Hall coefficient $R_H$ vs. $V_g$ for this sample, taken together with the $\rho_{xx}$ under the same temperatures.